\documentclass[a4paper,12pt]{article}
\usepackage{epsfig,amssymb,amsmath,textcomp,caption,subfigure}
\usepackage{graphicx}
\usepackage{amsfonts}
\usepackage[margin=1.0in]{geometry}
\usepackage[font={small}]{caption}
\usepackage[justification=centering]{caption}
\usepackage{amsmath}
\DeclareMathOperator{\Tr}{Tr}
\numberwithin{equation}{section}

\begin{document}
%

\newcommand{\be}{\begin{equation}}
\newcommand{\ee}{\end{equation}}
\newcommand{\bea}{\begin{eqnarray}}
\newcommand{\eea}{\end{eqnarray}}
\newcommand{\bean}{\begin{eqnarray*}}
\newcommand{\eean}{\end{eqnarray*}}
\font\upright=cmu10 scaled\magstep1
\font\sans=cmss12
\newcommand{\ssf}{\sans}
\newcommand{\stroke}{\vrule height8pt width0.4pt depth-0.1pt}
\newcommand{\Z}{\hbox{\upright\rlap{\ssf Z}\kern 2.7pt {\ssf Z}}}
\newcommand{\ZZ}{\Z\hskip -10pt \Z_2}
\newcommand{\C}{{\rlap{\upright\rlap{C}\kern 3.8pt\stroke}\phantom{C}}}
\newcommand{\R}{\hbox{\upright\rlap{I}\kern 1.7pt R}}
\newcommand{\HH}{\hbox{\upright\rlap{I}\kern 1.7pt H}}
\newcommand{\CP}{\hbox{\C{\upright\rlap{I}\kern 1.5pt P}}}
\newcommand{\identity}{{\upright\rlap{1}\kern 2.0pt 1}}
\newcommand{\half}{\frac{1}{2}}
\newcommand{\quart}{\frac{1}{4}}
\newcommand{\pr}{\partial}
\newcommand{\bm}{\boldmath}
\newcommand{\I}{{\cal I}} 
\newcommand{\M}{{\cal M}}
\newcommand{\N}{{\cal N}}
\newcommand{\e}{\varepsilon}
\newcommand{\ep}{\epsilon}
\newcommand{\bep}{\mbox{\boldmath $\varepsilon$}}
\newcommand{\Oh}{{\rm O}}
\newcommand{\bfk}{{\bf k}}
\newcommand{\n}{{\bf n}}
\newcommand{\q}{{\bf q}}
\newcommand{\x}{{\bf x}}
\newcommand{\y}{{\bf y}}
\newcommand{\w}{{\bf w}}
\newcommand{\X}{{\bf X}}
\newcommand{\Y}{{\bf Y}}
\newcommand{\z}{{\bar z}}
\newcommand{\tT}{{\tilde T}}
\newcommand{\tX}{{\tilde\X}}
\newcommand{\aveI}{{\langle I \rangle}}

\thispagestyle{empty}
\rightline{DAMTP-2015-57}
\vskip 5em
\begin{center}
{\bf \Large Electron Scattering Intensities and Patterson Functions of Skyrmions} 
\\[15mm]
{\bf \Large M. Karliner$^*$\footnote{email: marek@post.tau.ac.il}},
{\bf \Large C. King$^\dagger$\footnote{email: ck402@damtp.cam.ac.uk}} {\bf \Large and}
{\bf \Large N.S. Manton$^\dagger$\footnote{email: N.S.Manton@damtp.cam.ac.uk}} \\[10pt]

\vskip 3em
{$^*$\it 
School of Physics and Astronomy,\\
Raymond and Beverly Sackler Faculty of Exact Sciences,\\
Tel Aviv University, Tel Aviv 69978, Israel}

\vskip 2em
{$^\dagger$\it 
Department of Applied Mathematics and Theoretical Physics,\\
University of Cambridge, \\
Wilberforce Road, Cambridge CB3 0WA, U.K.}
\vspace{20mm}

\abstract{}

The scattering of electrons off nuclei is one of the best methods of
probing nuclear structure. In this paper we focus on electron scattering 
off nuclei with spin and isospin zero within the Skyrme model. 
We consider two distinct methods and simplify our calculations by use 
of the Born approximation. The first method is to calculate the form
factor of the spherically averaged Skyrmion charge density; the second uses 
the Patterson function to calculate the scattering intensity off randomly 
oriented Skyrmions, and spherically averages at the end. We compare
our findings with experimental scattering data. We also find 
approximate analytical formulae for the first zero and first
stationary point of a form factor.
\end{center}

\vskip 90pt
\leftline{Keywords:} 
\leftline{Skyrmion, Electron Scattering, Patterson Function} 

\vskip 5pt

\vfill
\newpage
\setcounter{page}{1}
\renewcommand{\thefootnote}{\arabic{footnote}}


\section{Introduction} 
\vspace{3mm}
The Skyrme model is a nonlinear field theory for nuclear physics that admits soliton solutions called Skyrmions \cite{MS}.
The Skyrme Lagrangian is
\be
L = \int \left\{ -\frac{F_{\pi}^2}{16} \Tr(R_{\mu}R^{\mu}) +
  \frac{1}{32e^2} \Tr([R_{\mu},R_{\nu}][R^{\mu},R^{\nu}])
+\frac{m^2_{\pi}F_{\pi}^2}{8} \Tr(U-I_2) \right\} d^3 \x \,,
\label{L}
\ee
where $U$ is the $SU(2)$-valued Skyrme field and $R_{\mu}= \partial_{\mu}UU^{\dag}$ is the right current.
The free parameters $F_{\pi},e$ and $m_{\pi}$ allow a choice of length and energy scale; they are absorbed by setting $\frac{F_{\pi}}{4e}$
as the Skyrme mass unit and $\frac{2}{eF_{\pi}}$ as the Skyrme length unit.
This leaves a rescaled pion mass $m=\frac{2}{eF_{\pi}}m_{\pi}$ which will be set to unity for all Skyrmions in this paper.
Skyrme field configurations have a topological charge $B$, the baryon
number, which takes values in the integers. $B$ is given by the
integral formula
\be
B=\int B_0(\x) \, d^3 \x \,, 
\ee
where
\be
B_{\mu}(\x)=\frac{1}{24\pi^2}\epsilon_{\mu \nu \alpha \beta} 
\Tr (\partial^{\nu}U U^{\dagger} \partial^{\alpha}U U^{\dagger} \partial^{\beta}U U^{\dagger})
\ee 
is the topological baryon current.

We are interested in electron scattering from a sample of many identical, uncorrelated nuclei. 
We model each nucleus as a quantised Skyrmion.
An analogous approach to pion-nucleon scattering reproduces experimental
phase shifts quite well \cite{Mattis:1984dh,Karliner:1986wq,Karliner:1986ap}.
The Skyrmion wavefunction gives the amplitude for the different orientations of the classical Skyrmion. 
If the spin is non-zero, then the wavefunction is not uniform with respect to body-fixed axes (and must satisfy 
Finkelstein--Rubinstein constraints). 
However, the projection of spin on to space-fixed axes is unconstrained if the nuclei are not polarised, so these 
projections occur with equal probability. 
The net effect is that all orientations of a Skyrmion occur with equal probability whatever the spin state. 

In this paper we will require each Skyrmion to have spin and isospin zero. 
These Skyrmions do not have an electric quadrupole or magnetic dipole moment; as a result their charge 
density is proportional to half their baryon density, $\rho(\x) =
\frac{1}{2}B_0(\x)$ \cite{ANW, CW}, which simplifies the calculation of scattering
intensities.
Note that only the $B=1$ Skyrmion has a spherically symmetric baryon density.

An isospin zero nucleus must have equal numbers of protons and neutrons. 
Within the nucleus, nucleons have a strong tendency to pair with like nucleons such that their spins are anti-parallel; 
therefore in an even-even nucleus all protons and neutrons pair in this manner, resulting in a spin and isospin zero nucleus. 
In an odd-odd nucleus we are left with a single proton and neutron after pairing; these have a stronger nuclear attraction 
between them if their spins are aligned, resulting in at least a spin one nucleus.
Therefore spin and isospin zero nuclei have baryon number $B=4N$ where $N$ is an integer.

The stationary Schr\"odinger equation for the electron involves the electrostatic potential $V$. 
In turn $V$ is related to the nucleus' charge density by Poisson's
equation (ignoring constants)
\be
\nabla^2 V = -\rho \,.
\label{Vrhorelation}
\ee
Phenomenological models of the nucleus describe the charge density as being spherically symmetric and approximated
quite accurately by a Fermi distribution $\rho (r) = \frac{\rho_0}{1+\exp{\frac{r-a}{c}}}$.
Variants with small oscillations and a central depression are also used, particularly for larger nuclei \cite{H}.
However, these variants require more parameters to fit the data closely and do not have a theoretical grounding
which explains their values.
We wish to see whether the Skyrme model can provide as good a fit with experimental scattering data, whilst being able
to vary only one parameter, the length scale (we use the same pion mass throughout this paper).

One method of calculating the electron scattering intensity off a
Skyrmion is to consider the semiclassical, spin zero quantum state,
and then calculate the form factor. The 
electron effectively scatters off the spherically averaged charge density of the Skyrmion.
We call this the quantum averaging method; it has been considered 
previously \cite{BC,MW}.

An alternative method of calculation is to consider the electrons as
moving fast and scattering off the Skyrmion in a brief moment. 
In this scattering time the Skyrmion has no time to change orientation. 
Nuclear rotational motion, from a classical perspective, is slow, 
like molecular rotational motion. We can therefore model the nucleus 
by a Skyrmion with a fixed orientation at the time that 
the electron scatters off it. We then average
over Skyrmion orientations, because the nuclei are not polarised
and all orientations are equally likely. We call this the classical 
averaging method.
 
For both methods of calculation we use the Born approximation
\cite{LL}, used routinely in quantum mechanical and X-ray scattering 
calculations.
The classical averaging method involves the Patterson function of the 
Skyrmion because the method is analogous to an X-ray powder diffraction 
calculation, where the crystal fragments have random orientations \cite{Gui}.

The electron is energetic and fast because we want to probe length
scales comparable to the nuclear radius.
In fact the electron can have energy of order 1 GeV and be
relativistic; therefore we use QED to calculate the cross section, 
resulting in the Mott scattering formula.
The Born approximation is then equivalent to the 
one-photon exchange approximation.

\section{Form Factors and Intensities} 
\vspace{3mm}

\subsection{Scattering from a fixed Skyrmion}
\vspace{2mm}
 
The expression for the electromagnetic current of a Skyrmion is
\be
\mathcal{J}_{\mu}=\frac{1}{2} B_{\mu} + I_{\mu}^3 \, ,
\ee
where $B_{\mu}$ is the baryon current and $I_{\mu}^3$ is the third
component of the isospin Noether current.
A Skyrmion in an isospin zero state has $I_{0}^3$ equal to zero and so
$\mathcal{J}_0$ is half the baryon density; 
thus we consider scattering off a charge density 
$\rho(\x) = \frac{1}{2} B_0(\x)$.

In the Born approximation, the electron scattering amplitude off a
Skyrmion with fixed orientation is a constant multiple of
\be
\widetilde V(\q) = \int V(\x) \, e^{-i\q \cdot \x} \, d^3 \x \,.
\ee

\begin{figure}[h]
\centering
\includegraphics[scale=1,keepaspectratio=true]{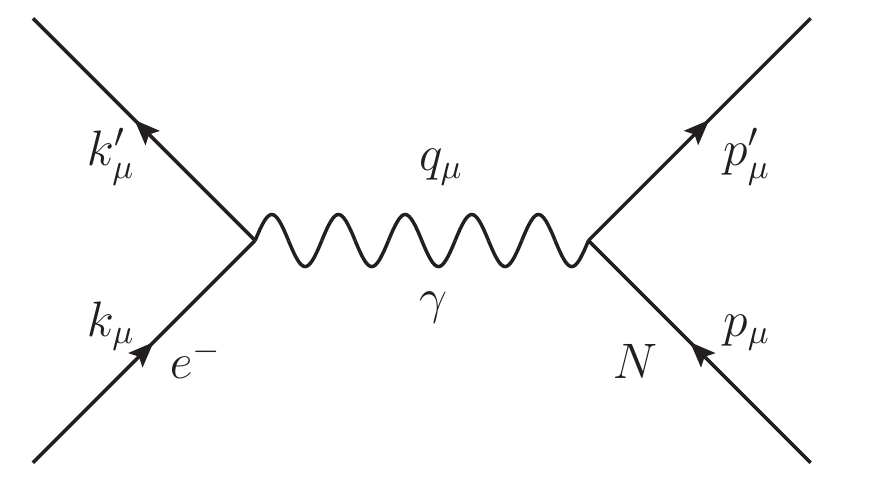}
  \captionof{figure}{Feynman diagram of the scattering process}
  \label{Feyn}
\end{figure}

Here $q^{\mu} = k^{\mu} - k'^{\mu}=(q^0,\q)$ is the momentum of the photon, where $k^{\mu} = (E,0,0,E)$ is the 
momentum of the incoming electron, and $k'^{\mu} = (E',E'\sin\theta \cos\phi,E'\sin\theta
\sin\phi,E'\cos\theta)$ is the momentum of an electron scattered in the
direction $(\theta, \phi)$ (Figure \ref{Feyn}).
We neglect the electron's mass because it is negligible compared to its kinetic energy.
The magnitude of $\q$ is $q = \frac{2 E \sin{\frac{\theta}{2}}}{\sqrt{1+\frac{2E}{M} \sin{\frac{\theta}{2}}}}$, 
where $M$ is the mass 
of the nucleus being scattered off. 
$V(\x)$ is the electrostatic potential of the Skyrmion and 
$\widetilde V$ is simply 
the Fourier transform of this. The differential cross section 
$\frac{d\sigma}{d\Omega}$ is proportional to $|\widetilde V(\q)|^2$.

It is desirable to have an expression for the scattering amplitude in 
terms of the charge density $\rho$ rather than the potential $V$, as it is the 
charge density (half the baryon density) that is computed for
Skyrmions. $V$ and $\rho$ are related by (\ref{Vrhorelation}), and 
fortunately, the Laplacian is simple in Fourier space. 
The Fourier transform of the charge density 
\be
F(\q) = \int \rho(\x) \, e^{-i\q \cdot \x} \, d^3 \x \,,
\label{F}
\ee
is called the form factor. Equation (\ref{Vrhorelation}) implies that 
$q^2 \widetilde V(\q) = F(\q)$, and the differential cross section \cite{DW} is
\be
\frac{d\sigma}{d\Omega} = \frac{(B \alpha)^2}{16 E^2 \sin^4{\frac{\theta}{2}}} \frac{\cos^2{\frac{\theta}{2}}}{1+\frac{2E}{M} \sin^2{\frac{\theta}{2}}}|F(\q)|^2.
\ee
Here $\alpha$ is the fine-structure constant and $B$ is the Baryon number.
The prefactor is present for any charge density, so it is
generally the modulus of the form factor $|F(\q)|$ that is
discussed. The scattering intensity is defined as $|F(\q)|^2$.

\subsection{The Quantum Averaged Intensity}
\vspace{2mm}

We assume now that electrons scatter off quantised Skyrmions in 
their spin zero ground state. The electrons therefore scatter off 
the spherically averaged charge density. This is obtained by integrating 
over a rotation matrix $R$, depending on three Euler 
angles $(\alpha, \beta, \gamma)$, with the standard normalised measure
\be
dR = \frac{1}{8\pi^2} \sin\beta \, d\alpha \, d\beta \, d\gamma \,.
\ee
The spherically averaged charge density is
\be
\rho(r) = \int \rho(R\x) \, dR
\ee
and is just a function of the radial coordinate $r$.
The form factor of the spherically averaged charge density is then
\be
\mathcal{F}(q^2)= \int \rho(r) \, e^{-i\q \cdot \x} \, d^3 \x \,,
\ee
and only depends on $q^2$.
(A calligraphic letter corresponds to a spherically averaged roman letter.)

To simplify this integral we may assume that $\q = q(0,0,1)$. Then
using polar coordinates,
\bea
\mathcal{F}(q^2) &=& \int \rho(r) \, 
e^{-iqr \cos\theta} r^2 \sin\theta \, dr d\theta d\phi \nonumber \\ 
&=& 2\pi\int \rho(r) \, e^{-iqr \cos\theta} r^2 \sin\theta \,
dr d\theta \nonumber \\
&=& 2\pi\int \rho(r) \, \left[\frac{e^{-iqr \cos\theta}}{iqr}
\right]_0^{\pi} r^2 \, dr \nonumber \\
&=& 4\pi \int_0^{\infty} \rho(r)\frac{\sin(qr)}{qr} r^2 \, dr 
\equiv 4\pi \int_0^{\infty} \rho(r) j_0(qr) r^2 \, dr \,,
\label{spherformfac}
\eea
where $j_0(u) = \frac{\sin u}{u}$ is the zeroth spherical Bessel
function. 

As the Fourier transform is a linear operation, and the measure
$d^3\x$ is rotationally invariant, the form factor of the
spherically averaged charge density is the same as the spherically
averaged form factor of the initial charge density. Therefore an
alternative expression for $\mathcal{F}(q^2)$ is 
\be
\mathcal{F}(q^2) = \int F(R\q) \, dR \,,
\ee
where $F(\q)$ is the form factor (\ref{F}). Although this expression
is less explicit than (\ref{spherformfac}), we will have some use for it.

This form factor (or rather its modulus $|\mathcal{F}|$) is the
function that is usually extracted from the experimental cross section
data and gives one information about the spherically
averaged charge density. 

\subsection{The Angular Velocity of a Skyrmion}
\vspace{2mm}
It is important to know which electron energies allow us to treat the nucleus as having a fixed orientation, and therefore
allow us to use the classical approximation.

Whilst the Skyrmion is originally spin $0$ and therefore stationary, the electron scattering could cause the Skyrmion to 
rotate. It is difficult to know exactly what the resulting angular velocity would be, but we can consider the angular velocity of a 
spin $1$ Skyrmion to get an estimate.

In order to calculate the angular velocity of the $B=4$ Skyrmion, we need to choose values for $F_{\pi}$ and $e$.
The best way to do this is to calibrate to the mass and mean charge radius of an alpha particle, which requires 
$F_{\pi}=87.3$ MeV and $e=3.65$. 
This gives an energy unit of $\frac{F_{\pi}}{4 e}=5.98$ MeV and a length unit of $\frac{2}{e F_{\pi}}=1.24 \, \rm{fm}$. 
Thus the Skyrme unit of energy length becomes $\frac{F_{\pi}}{4 e} \, \frac{2}{e F_{\pi}}=7.39$ MeV fm, whilst $\hbar = 197$ MeV fm;
therefore $\hbar=26.7=2e^2$ in Skyrme units. 

Consider the $B=4$ Skyrmion rotating about the 3rd axis with angular momentum $L_3=1$; 
this means that the angular velocity 
$\omega=\frac{\hbar}{V_{33}}$, where $V_{33}=663$ is the $(3,3)$ component of the spin inertia tensor in Skyrme units.
Using this angular velocity and the mean charge radius (around $1.36$ in Skyrme units), we calculate that the speed at the surface of the Skyrmion is 
around $\frac{1}{20}$ of the speed of light. 
This means that the time for the electron (travelling close to the speed of light) to cross the nucleus is around 
$\frac{1}{125}$ of the time for the nucleus to complete a full rotation.

For a roughly spherical object the moment of inertia is proportional to $MR^2$, where $M$ is the mass of the object and $R$ 
is the radius, with the mass being proportional to $R^3$. 
Thus, as we increase the baryon number of the Skyrmion (thereby increasing its mean charge radius), its angular
velocity has a large supression due to the increased size.
This means that the larger Skyrmion rotates more slowly than the $B=4$ Skyrmion.

Therefore for all the Skyrmions considered, the time for the electron to scatter off the nucleus is much less than the period
for a full rotation of the nucleus. Thus it is reasonable to treat the nucleus as having a fixed orientation during the
scattering process for all electron energies of interest.

\subsection{The Classically Averaged Intensity}
\vspace{2mm}

Here we assume that each electron scatters off a Skyrmion with fixed
but random orientation, the Skyrmion having a non-spherically symmetric charge density. 
The contributions of the different Skyrmions add incoherently, so the scattering intensities 
need to be spherically averaged over the orientations to find the total intensity. 

For a Skyrmion with fixed orientation, the intensity is
\bea
I(\q) &=& |F(\q)|^2 \label{|V|^2} \\
&=& \int \rho(\x) \, \rho(\x') \, e^{i\q \cdot (\x - \x')} \, d^3 \x \,
d^3 \x' \,.
\label{VV}
\eea
If one writes $\x' = \x + \w$, and replaces the $\x'$ integral by an integral over $\w$, then
\be
I(\q) = \int \rho(\x) \, \rho(\x + \w) \, e^{-i\q \cdot \w} \, d^3 \x
\, d^3 \w \,.
\ee
The $\x$ integral here,
\be
P(\w) = \int \rho(\x) \, \rho(\x + \w) \, d^3 \x \,,
\ee
is called the Patterson function. 
It is an autocorrelation function, but not a convolution. 
Then
\be
I(\q) = \int P(\w) \, e^{-i\q \cdot \w} \, d^3 \w \,,
\ee
so the intensity is the Fourier transform of the Patterson function \cite{Gui}.

Changing the orientation involves a rotation matrix $R$ depending on
Euler angles $(\alpha, \beta, \gamma)$ as before. We can define a
rotationally averaged Patterson function
\be
P(w) = \int P(R\w) \, dR \,,
\ee
which is just a function of $w = |\w|$. Then, by repeating the steps 
in subsection $2.2$, we obtain the classically averaged 
intensity
\be
\mathcal{I}(q^2) = 4\pi \int_0^{\infty} P(w) j_0(qw) w^2 \, dw \,.
\ee
Again there is an alternative expression, directly in terms of the
charge density,
\be
\mathcal{I}(q^2) = \int \rho(\x) \, \rho(\x') \, e^{iR\q \cdot (\x -
\x')} \, d^3 \x \, d^3 \x' \, dR \,,
\label{PatFT}
\ee
which simplifies to
\be
\mathcal{I}(q^2) = \int \rho(\x) \, \rho(\x') j_0(q|\x - \x'|) 
\, d^3 \x \, d^3 \x' \,,
\ee
a Debye-type scattering formula.

Note that we can also express this classically averaged intensity 
in terms of the form factor. From (\ref{PatFT}),
\begin{equation}
\mathcal{I}(q^2)=\int |F(R\q)|^2 \, dR \,.
\end{equation}

While it is straightforward to calculate $\mathcal{I}(q^2)$ from a
given charge density, one cannot reconstruct
the charge density from $\mathcal{I}(q^2)$. 
There are essential ambiguities in the reconstruction.

\subsubsection{Properties of Patterson Functions}
\vspace{2mm}

The Patterson function has a couple of properties which hold 
for any charge distribution. It is invariant under $\w \rightarrow -\w$; 
this follows easily from the definition:
\be
P(\w) = \int \rho(\x) \, \rho(\x + \w) \, d^3 \x \, = \int \rho(\x - \w) \, \rho(\x) \, d^3 \x \, = P(-\w) \,,
\ee
where the second equality follows from a simple change of variables.

The second universal property is that $P(\w)$ is maximal at
$\w=\textbf{0}$. We have, for any real $a$,
\be
\int \bigl( \rho(\x) \, + a \rho(\x + \w) \bigr)^2 \, d^3 \x \geq 0 \,,
\ee
and after expanding the bracket,
\be
\int \rho(\x)^2 \, d^3 \x +  2 a \int \rho(\x) \, \rho(\x + \w) 
\, d^3 \x + a^2 \int \rho(\x + \w)^2 \, d^3 \x \geq 0 \,.
\ee
The first and third integrals are both equal to 
$P(\textbf{0})= \int \rho(\x)^2 \, d^3 \x$, because the domain of integration 
is all of space. The second integral is equal to $P(\w)$, therefore
\be
(1 + a^2)P(\textbf{0}) + 2aP(\w) \geq 0
\ee
for all $a$. This implies the discriminant condition
\be
P(\w)^2 \leq P(\textbf{0})^2 \,,
\ee
so the Patterson function is maximal at the origin.

\subsection{Comparison of the Scattering Intensities}
\vspace{2mm}

The classically averaged intensity $\mathcal{I}(q^2)$ is 
the average of $|F(\q)|^2$ over all directions $\q$,
\begin{equation}
\mathcal{I}(q^2) =\int |F(R\q)|^2 \, dR \,,
\label{IaveF}
\end{equation}
whereas the quantum averaged intensity $|\mathcal{F}(q^2)|^2$ is the
modulus squared of the average of $F(\q)$ over all directions $\q$,
\begin{equation}
|\mathcal{F}(q^2)|^2 = \left|\int F(R\q) \, dR \right|^2 .
\end{equation}
These are fundamentally different objects, and a Cauchy-Schwartz inequality informs us that 
$|\mathcal{F}(q^2)|^2 \leq \mathcal{I}(q^2)$.

We will present $|\mathcal{F}|$ and $\sqrt{\mathcal{I}}$ against 
$q^2$ for all the Skyrmions in this paper, 
in order to illustrate the differences between the quantum and 
classically averaged intensities. Experimental electron scattering data 
is usually presented as the modulus of the form factor 
$|\mathcal{F}|$ plotted against $q^2$.
It should be noted that extrema of $\mathcal{F}$ appear as extrema of
$|\mathcal{F}|$, but $|\mathcal{F}|$ has additional sharp minima at
the zeros of $\mathcal{F}$. 
Thus it can be helpful to plot $\mathcal{F}$ as well as
$|\mathcal{F}|$, but this is only known in theoretical calculations,
and not from the experimental data.

The formula (\ref{IaveF}) shows that $\mathcal{I}(q^2)$ is
non-negative, and $\mathcal{I}(q^2)$ is zero for some value of $q^2$
only if $F(\q)=0$ for all $\q$ with $q^2 = |\q|^2$.
For a non-spherically symmetric charge distribution 
this is a shell of independent conditions, not generally all
satisfied, so we would not expect $\mathcal{I}(q^2)$ to have any zeros.

For a spherically symmetric charge distribution, $F(\q)$ is independent
of the direction of $\q$, so this shell of conditions collapses
to a single condition, and the intensity
$\mathcal{I}(q^2)$ will generically have zeros. 
There are also extreme examples of non-spherically symmetric charge 
distributions for which $\mathcal{I}(q^2)$ has zeros.

This shows that there is an important difference between the electron
scattering intensities of spherically and non-spherically symmetric models
of nuclei; we expect some zeros in the former case and no zeros in the latter. 
This is also a difference between quantum and classically averaged intensities.
Unfortunately, it is a difference that is difficult to probe by experiment. 
In an electron scattering experiment the differential cross section 
can only be found for a discrete set of momenta, $\q$. 
This means that there is effectively no chance of finding a value of $q^2$
that yields a zero in the intensity; thus, we are not 
able to distinguish between a sharp minimum and a zero.

\section{Scattering Intensities of $B=4N$ Skyrmions}
\vspace{3mm}

\subsection{Calibration of Skyrme Units} 
\vspace{2mm}

In order to compare our intensities with experimental scattering data, we need to perform some calibrations. 
We normalise the charge density $\rho$, so that it integrates to $1$ over all space. 
This has the effect of making $\sqrt{\mathcal{I}}$ and $|\mathcal{F}|$ have value $1$ at $q^2=0$. 
In graphs of experimental scattering data the form factor is usually presented as $|\mathcal{F}(q^2)|/|\mathcal{F}(0)|$, so our choice of 
normalisation enables easier comparison with the data.

The Skyrmion's length scale is measured in Skyrme units, and the parameters $F_{\pi}$ and $e$ of the model (\ref{L}) decide the 
conversion factor from Skyrme units to MeV and fermi. 
We must decide on a length scale of the nucleus to calibrate to, and there are a few choices. 
The most sensible choice \cite{MW2} appears to be the (root mean square) charge
radius, the square root of
\begin{equation}
\langle r^2 \rangle=\int \rho(\x) \, r^2 \, d^3 \x = 4\pi
\int_0^{\infty} \rho(r) \, r^4 \, dr \,.
\end{equation}
This is a measure of the total size of a nucleus, and $\langle r^2
\rangle$ appears naturally in the spherically averaged form factor. 
For small $q^2$
\begin{equation}
\mathcal{F}(q^2) = 4\pi \int_0^{\infty} \rho(r)\frac{\sin qr}{qr} r^2 \, dr 
\approx 4\pi \int_0^{\infty} \rho(r)\left(1-\frac{1}{3!}q^2r^2\right) r^2
\, dr = 1-\frac{q^2}{6}\langle r^2 \rangle \,.
\end{equation}
We see that $\langle r^2 \rangle$ is $-6$ times the gradient of 
$\mathcal{F}(q^2)$ with respect to $q^2$ at the 
origin, so by calibrating to this we will have the correct charge radius.

An alternative calibration would be to fix the length scale such that the first minimum of the calculated
and experimental scattering intensities are at the same value of $q^2$. 
But this seems less fundamental, especially since the relation between characteristic length scales and locations of minima 
in momentum space is a complicated one.

The parameter set $F_{\pi}=75.2$ MeV, $e=3.26$ and $m_{\pi}=138$ MeV provide reasonable accuracy for the masses and mean 
charge radii of several Skyrmions \cite{MW2}. However, in this paper we decided to recalibrate the parameter values such
that the mean charge radius was correct for all the Skyrmions considered and the associated parameter $m$ was fixed at $1$.
Therefore the parameters depend on baryon number.

Experimental values in Table \ref{T2} for $\langle r^2 \rangle^\frac{1}{2}$ are taken from \cite{AM}. The charge radius of Beryllium-8 is unknown
due to its instability, so we will use the charge radius of Beryllium-9 instead (the instability also results in a lack
of experimental scattering data to compare with, thus calibration is less important). $B=108$ also creates a problem; an 
isotope with equal numbers of protons and neutrons at that baryon number would be highly unstable. In addition, charge
radii calculations are usually model dependent in this region; we choose $\rm{Ag}_{47}^{107}$ as a candidate, taking the
value of $\langle r^2 \rangle$ from \cite{LB}.

\begin{table} [h]
\centering
    \begin{tabular}{| c | c | c | c | c | c | c |}
    \hline
    Baryon Number & $B=4$ & $B=8$ & $B=12$ & $B=16$ & $B=32$ & $B=108$   \\ \hline
    $\langle r^2 \rangle^\frac{1}{2} \rm{fm} $ & 1.68  & 2.52 & 2.47 & 2.70 & 3.26 & 4.50 \\ \hline
    \end{tabular}
    \caption{Table of charge radii for nuclei with zero
      isospin. The entries for $B=8$ and $B=108$ are estimates (see text).}
    \label{T2}
    \end{table}  

We ignore most of the Skyrmions after $B=16$ because there is less confidence in their baryon densities.
There is more confidence in $B=32$ and $B=108$  because of their cubic structure.

\subsection{$B=4$}
\vspace{2mm}

\begin{figure}[h]
\centering
\begin{minipage}{.5\textwidth}
  \centering
\includegraphics[scale=0.5,keepaspectratio=true]{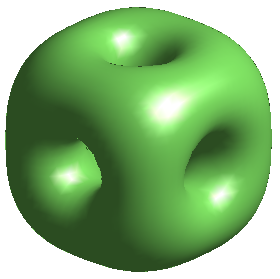}
  \captionof{figure}{Charge Density Isosurface\\ of the $B=4$ Skyrmion}
  \label{CD4}
\end{minipage}%
\begin{minipage}{.5\textwidth}
  \centering
\includegraphics[scale=0.5,keepaspectratio=true]{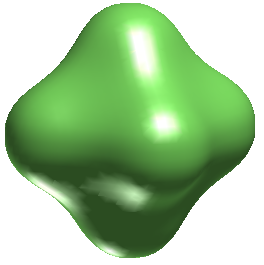}
  \captionof{figure}{Patterson Function Isosurface\\ of the $B=4$ Skyrmion}
  \label{PF4}
\end{minipage}
\end{figure}

The $B=4$ Skyrmion is cubically symmetric (Figure \ref{CD4}). 
The charge density (half the baryon density) is easily evaluated, and from this the spherically averaged
form factor is calculated. Evaluating the Patterson function leads to
the classically averaged intensity.

\subsubsection{Patterson Function of the $B=4$ Skyrmion}
\vspace{1mm}
There are two main ways to see how the Patterson function $P(\w)$ varies with shift vector $\w$.
The first is to take an isosurface plot of $P(\w)$, which shows surfaces of constant $P$. 
The second is to take a planar slice in $\w$-space and plot contours of constant $P$. 
All representations of the Patterson function will be in Skyrme units; these will then be converted into fermi (fm) to 
calculate intensities.

The Patterson function of the $B=4$ Skyrmion is spherically symmetric
for small $w$, where $w = |\w|$, and as $w$ 
increases it takes on the shape of a concave octahedron (Figure \ref{PF4}). 
The slowest descent occurs along shift vectors $\w$ corresponding to the primary $x,y,z$ axes of the $B=4$ Skyrmion. 
We see from Figure \ref{CD4} that the regions of highest charge density are the corners and edges of the $B=4$ cube, so when
we shift along the Cartesian axes by $w = a \approx 1.4$, the distance
between opposite faces, there is strong overlap of high density regions.
This is in contrast to shifting along the body diagonal direction; here one of the corners travels into the hollow
centre of the cube, resulting in a weak overlap.
As $w$ increases further, the concave edges of the octahedron become convex. 
This is followed by the faces becoming convex, producing approximately cubic contours.

\begin{figure}[h]
\centering
\begin{minipage}{.5\textwidth}
  \centering
\includegraphics[scale=0.55,keepaspectratio=true]{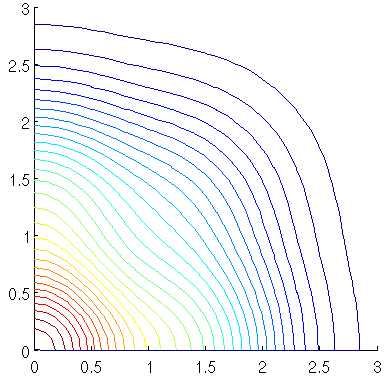}
  \captionof{figure}{Patterson Function Contours in\\ the $(x,y)$-plane for the $B=4$ Skyrmion}
  \label{PC4}
\end{minipage}%
\begin{minipage}{.5\textwidth}
\centering
 \includegraphics[scale=0.55]{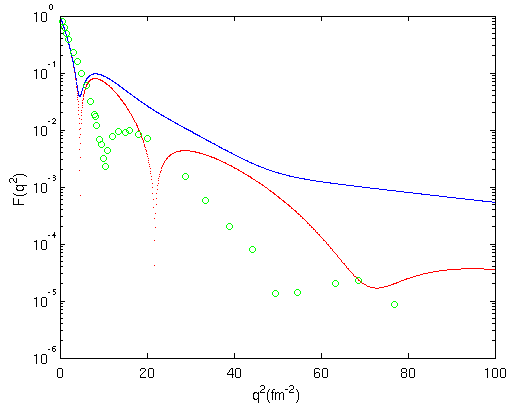}
  \captionof{figure}{Comparison of $\sqrt{\mathcal{I}}$ (blue) and
    $|\mathcal{F}|$ (red) of the $B=4$ Skyrmion with the Form Factor from 
  Experimental Scattering Data (green) for the $\rm{He}^4$ Nucleus}
\label{EXP4}
\end{minipage}
\end{figure}

Figure \ref{PC4} shows a contour plot of $P(\w)$ in the $(x,y)$-plane. 
Red corresponds to the highest values of $P$ and blue to the lowest; this colour scheme is retained for all contour plots 
in this paper.
Taking a slice in the $(x,y)$-plane allows us to see in which regions $P$ varies the least. 
We observe that the contours are most widely spaced at the points $(0,a)$, $(a,0)$ and $(a,a)$ (ignoring $(0,0)$, which is 
always a maximum).
The Patterson function approximately levels off at shift vectors $\w$ corresponding to the three distinct 
types of separation vector between corners of the $B=4$ Skyrmion (along an edge, diagonally across a face, and diagonally
across the body).

\subsubsection{Intensity for the $B=4$ Skyrmion}
\vspace{1mm}
In Figure \ref{EXP4} the blue curve represents $\sqrt{\mathcal{I}}$ and the red curve represents $|\mathcal{F}|$ 
for the $B=4$ Skyrmion, plotted on a logarithmic scale; this colour scheme and log scale are retained throughout the paper. 
The Skyrmion has a characteristic length of around $1 \, \rm{fm}$, so
our main interest is in $q^2$ greater than $1 \, \rm{fm}^{-2}$. 
We observe a minimum at $q^2 \approx 4.5 \, \rm{fm}^{-2}$ and close agreement between $\sqrt{\mathcal{I}}$ and 
$|\mathcal{F}|$ for $q^2 <15 \, \rm{fm}^{-2}$.

\subsubsection{Comparison of Intensity with Experimental Data for the $B=4$ Skyrmion}
\vspace{1mm}

The $B=4$ Skyrmion is the best of the $B=4N$ Skyrmions to compare with experimental data because electron scattering experiments on 
$\rm{He}^4$ have been done up to $q^2 = 80 \, \rm{fm}^{-2}$. 
This is almost an order of magnitude greater than the highest energy data for other $B=4N$ nuclei.

In Figure \ref{EXP4} the green circles represent experimental
$|\mathcal{F}|$ data for electron scattering off a $\rm{He}^{4}$ nucleus
\cite{CKOS}.
The shapes of the theoretical and data curves are quite similar for low values of $q^2$, which 
is to be expected, because normalisation forces all three of them to have both $\mathcal{F}(0)=1$ and the same gradient at $q^2=0$. 
There is a quite a difference in the $q^2$ value at the first minimum and an order of magnitude discrepancy in 
$|\mathcal{F}_{\rm max}|$, the value of $|\mathcal{F}|$ at the first maximum beyond the first minimum. 
The value of $q^2$ at the first minimum could be fixed by a different length calibration (although the derivatives at the
origin would no longer match), but $|\mathcal{F}_{\rm max}|$ is much harder to rectify because it is independent of the
calibration. 

The experimental data curve has a second minimum at $q^2 \approx 52 \, \rm{fm}^{-2}$, whereas the classically averaged Skyrmion curve (blue)
fails to predict a second diffraction minimum at any value of $q^2$.
The quantum averaged Skyrmion curve (red) fares better, although it has too many minima in the region that we are considering. 
However, if the Skyrme length unit is recalibrated such that the first minima of the quantum averaged curve and data curve 
are at the same value of $q^2$, then the second minima of the curves are within a few $\rm{fm}^{-2}$ of each other, and 
the third minimum of the quantum averaged curve moves outside the region that has been probed by experiment.
This does not fix all of the problems; the values of $|\mathcal{F}|$ at the second maxima of the curves would disagree by
two orders of magnitude.

\subsection{$B=8$}
\vspace{2mm}

\begin{figure}[h]
\centering
\begin{minipage}{.5\textwidth}
  \centering
  \includegraphics[scale=0.45,keepaspectratio=true]{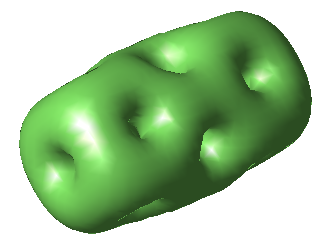}
  \captionof{figure}{Charge Density Isosurface\\ of the $B=8$ Skyrmion}
  \label{CD8T8L}
\end{minipage}%
\begin{minipage}{.5\textwidth}
  \centering
\includegraphics[scale=0.5,keepaspectratio=true]{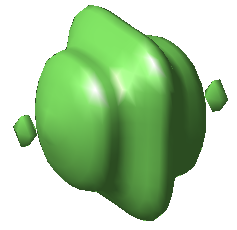}
  \captionof{figure}{Patterson Function Isosurface\\ of the $B=8$ Skyrmion}
  \label{PF8T}
\end{minipage}
\end{figure}

The $B=8$ Skyrmion is a pair of $B=4$ cubes joined on a face. It comes
in two different forms with very similar energies: linear and
twisted \cite{BMS}. There is only a small difference between the form factors and
Patterson functions of these two Skyrmions; therefore we 
will only present the twisted variant (Figure \ref{CD8T8L}) in this paper.

\subsubsection{Patterson Function of the $B=8$ Skyrmion}
\vspace{1mm}

\begin{figure}[h]
\centering
\begin{minipage}{.5\textwidth}
    \centering
\includegraphics[scale=0.55,keepaspectratio=true]{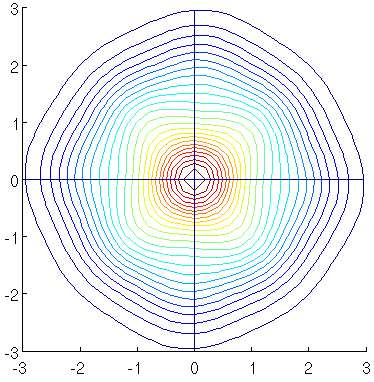}
  \captionof{figure}{Patterson Function Contours\\ in the $(x,z)$-plane for the $B=8$ Skyrmion}
  \label{PC8T2}
\end{minipage}%
\begin{minipage}{.5\textwidth}
  \centering
\includegraphics[scale=0.55,keepaspectratio=true]{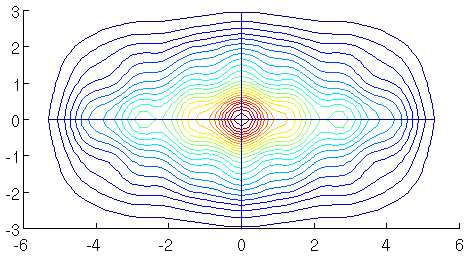}
  \captionof{figure}{Patterson Function Contours in\\ the $(x,y)$-plane for the  $B=8$ Skyrmion}
  \label{PC8T1}
\end{minipage}
\end{figure}

\noindent
The $B=8$ Skyrmion has $D_{4h}$ symmetry and this symmetry is inherited by the Patterson function (Figure \ref{PF8T}).
In the $(x,z)$-plane (Figure \ref{PC8T2}) (the $y$-axis being the $C_4$ axis) the Patterson function's contours 
display similar behavior to those of the $B=4$ Skyrmion, which is not surprising, because shifting in this plane does not allow 
the two $B=4$ cubes to overlap each other.
There is a slight reduction in concavity of the Patterson function for small $w$ in comparison with the $B=4$ 
Skyrmion, because the middle section of the $B=8$ Skyrmion has different intrinsic axes to the cubes either side of it.

We see different behavior to the $B=4$ case when shifts have a $y$-component. 
The isosurface has an oscillatory nature, with
level regions for $\w$ close to the various
separation vectors between the $16$ cube corners of the $B=8$ Skyrmion. 
There is a small range of values of $P$ which exhibit disjoint isosurfaces; two balls occur when the 
shift vector is in the $y$-direction so that one cube coincides
almost perfectly with the other cube. 
These balls indicate local maxima of the Patterson function, a feature not present for the $B=4$ Skyrmion.

A contour slice in the $(x,y)$-plane (Figure \ref{PC8T1}) confirms
these maxima. The slice also shows the locations of saddle points.\\

\subsubsection{Intensity for the $B=8$ Skyrmion}
\vspace{1mm}

\begin{figure}[h]
\centering
\includegraphics[scale=0.55,keepaspectratio=true]{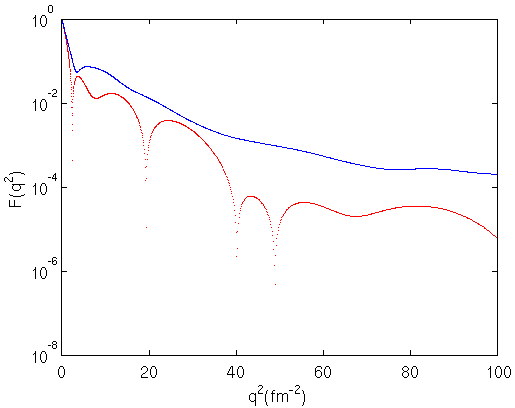}
  \captionof{figure}{$\sqrt{\mathcal{I}}$ and $|\mathcal{F}|$ of the $B=8$ Skyrmion}
  \label{FF8T}
\end{figure}

The first minimum of $|\mathcal{F}|$ is not very close to a stationary point of $\sqrt{\mathcal{I}}$ 
(Figure \ref{FF8T}).
The first minimum is related to the size of the Skyrmion, and we see that the intensity for the $B=8$ 
Skyrmion has a first minimum at a lower value of $q^2$ ($\approx 3.4 \, \rm{fm}^{-2}$) than the $B=4$ Skyrmion.
This is due to the $B=8$ Skyrmion having a large charge radius, caused
by its extension in the $y$-direction. 

The spin zero $B=8$ Skyrmion represents $\rm{Be}^8$, but this is very unstable, and there is no high-energy electron 
scattering data to compare with.

\subsection{$B=12$}
\vspace{2mm}

The $B=12$ Skyrmion comes in two different forms: triangular
and linear \cite{BMS}. The triangular Skyrmion, which has $D_{3h}$ symmetry, corresponds to the ground
state of $\rm{C}^{12}$ modeled as a triangle of alpha particles. The linear Skyrmion, 
which has $D_{4h}$ symmetry, corresponds to the Hoyle state modeled as a
chain of alpha particles \cite{LM}.

\vspace{1mm}
\begin{figure}[h]
\centering
\begin{minipage}{.5\textwidth}
  \centering
\includegraphics[scale=0.45,keepaspectratio=true]{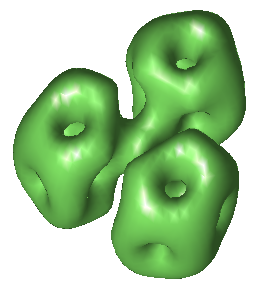}
  \captionof{figure}{Charge Density Isosurface\\ of the triangular $B=12$ Skyrmion}
  \label{CD12T}
\end{minipage}%
\begin{minipage}{.5\textwidth}
  \centering
\includegraphics[scale=0.5,keepaspectratio=true]{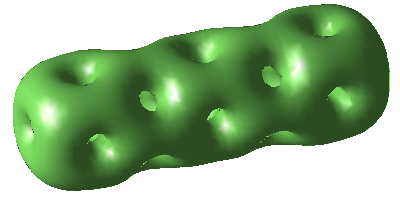}
  \captionof{figure}{Charge Density Isosurface\\ of the linear $B=12$ Skyrmion}
  \label{CD12L}
\end{minipage}
\end{figure}

\subsubsection{Patterson Functions of $B=12$ Skyrmions}

\begin{figure}[h]
\centering
\begin{minipage}{.5\textwidth}
  \centering
\includegraphics[scale=0.45,keepaspectratio=true]{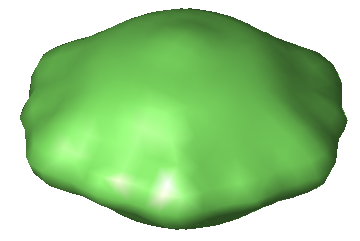}
  \captionof{figure}{Patterson Function Isosurface\\ of the triangular
    $B=12$ Skyrmion}
  \label{PF12T}
\end{minipage}%
\begin{minipage}{.5\textwidth}
  \centering
\includegraphics[scale=0.5,keepaspectratio=true]{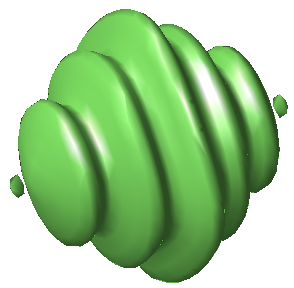}
  \captionof{figure}{Patterson Function Isosurface\\ of the linear $B=12$ Skyrmion}
  \label{PF12L}
\end{minipage}
\end{figure}

\begin{figure}[h]
\centering
\begin{minipage}{.5\textwidth}
  \centering
\includegraphics[scale=0.45,keepaspectratio=true]{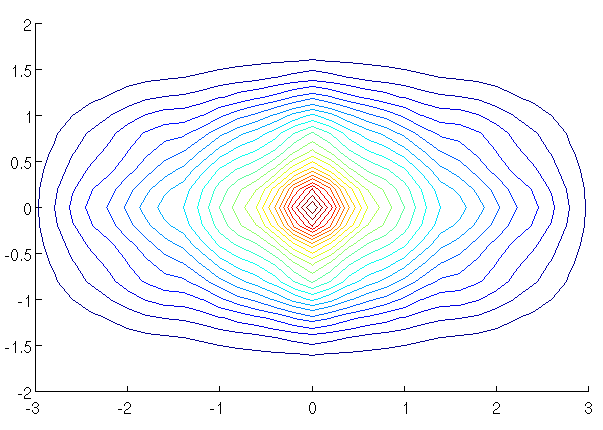}
  \captionof{figure}{Patterson Function Contours\\ in the $(x,z)$-plane for the triangular\\ $B=12$ Skyrmion}
  \label{PC12T1}
\end{minipage}%
\begin{minipage}{.5\textwidth}
  \centering
\includegraphics[scale=0.45,keepaspectratio=true]{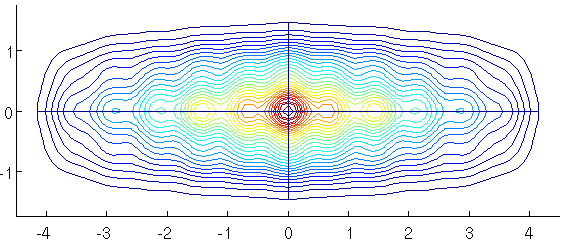}
  \captionof{figure}{Patterson Function Contours\\ in the $(x,y)$-plane for the linear\\ $B=12$ Skyrmion}
  \label{PC12L1}
\end{minipage}
\end{figure}

\underline{$B=12$ triangular:}\\
The triangular Skyrmion has $C_3$ symmetry about the $z$-axis (Figure \ref{CD12T}), and this combines with the intrinsic 
$\w \rightarrow -\w$ symmetry of the Patterson function to give an overall $D_{6h}$ symmetry of $P(\w)$. 
The isosurface plots resemble a rounded hexagonal bipyramid (Figure \ref{PF12T}).
The contour slices chosen for this Skyrmion are defined differently to
slices for other Skyrmions; the $(x,z)$-plane is chosen to include the 
origin and the centre of one of the three cubes. 
This plane is featured in Figure \ref{PC12T1}. 
We see that the axis directions are distinguished 
for all but the largest shift vectors. This is because these
directions correspond to one of the cubes
shifting along an intrinsic axis, which results in a strong overlap of charge density.
Rotating this plane by $30$\textdegree\ about the $z$-direction results in similar contours, but they are less oblong. 
When the cubes shift in the $z$-direction, they have no overlap with themselves after moving more than a side length,
so the Patterson function decreases rapidly.

\begin{figure}[h]
\centering
\begin{minipage}{.5\textwidth}
  \centering
\includegraphics[scale=0.55,keepaspectratio=true]{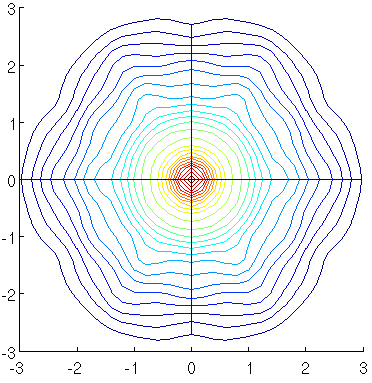}
  \captionof{figure}{Patterson Function Contours\\ in the $(x,y)$-plane for the triangular\\ $B=12$ Skyrmion}
  \label{PC12T2}
\end{minipage}%
\begin{minipage}{.5\textwidth}
  \centering
\includegraphics[scale=0.55,keepaspectratio=true]{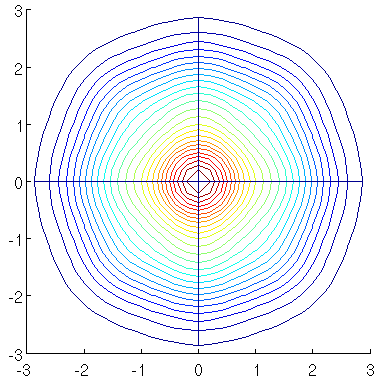}
  \captionof{figure}{Patterson Function Contours\\ in the $(x,z)$-plane for the linear\\ $B=12$ Skyrmion}
  \label{PC12L2}
\end{minipage}
\end{figure}

The contours in the $(x,y)$-plane (Figure \ref{PC12T2}) have a
hexagonal structure, which reflects the symmetry of the Skyrmion. \\

\noindent
\underline{$B=12$ linear:}\\
The $B=12$ linear Skyrmion is a chain of three $B=4$ cubes with $D_{4h}$
symmetry about the $y$-axis (Figure \ref{CD12L}), and we expect similar characteristics to the 
$B=8$ twisted Skyrmion. 
The Patterson function (Figure \ref{PF12L}) also shares characteristics with that of the $B=8$ Skyrmion.

In the $(x,y)$-plane (Figure \ref{PC12L1}) there is a series of maxima
and saddle points on the $y$-axis, with 
the similar oscillatory decreasing behavior that we saw previously. 
The presence of an extra cube increases the number of 
shift vectors between cube corners, which results in a 
greater number of local maxima and saddle points.

The Patterson function contours in the $(x,z)$-plane (Figure
\ref{PC12L2}) are almost the same as for the $B=8$ Skyrmion; the presence of three $B=4$ cubes 
rather than two makes little difference.

\subsubsection{Intensities of $B=12$ Skyrmions}
\vspace{1mm}

\begin{figure}[h]
\centering
\begin{minipage}{.5\textwidth}
  \centering
\includegraphics[scale=0.50,keepaspectratio=true]{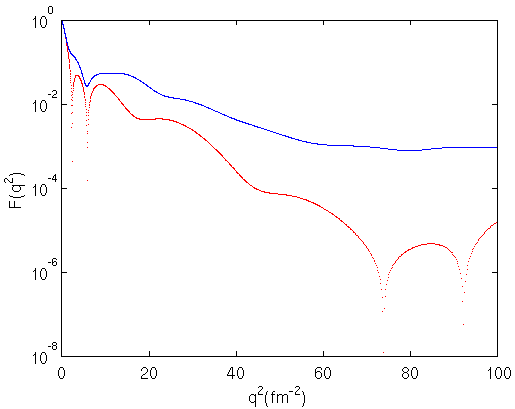}
  \captionof{figure}{$\sqrt{\mathcal{I}}$ and $|\mathcal{F}|$ for the\\ triangular $B=12$ Skyrmion}
  \label{FF12T}
\end{minipage}%
\begin{minipage}{.5\textwidth}
  \centering
\includegraphics[scale=0.50,keepaspectratio=true]{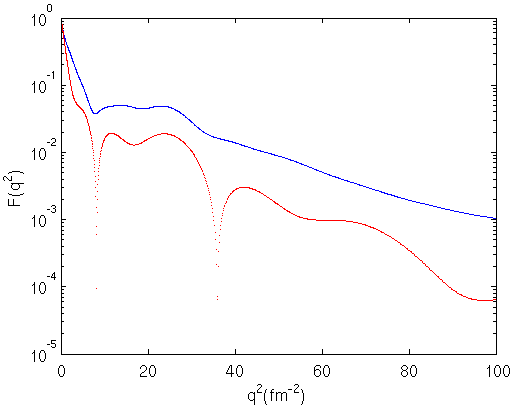}
  \captionof{figure}{$\sqrt{\mathcal{I}}$ and $|\mathcal{F}|$ for the\\ linear $B=12$ Skyrmion}
  \label{FF12L}
\end{minipage}
\end{figure}

\underline{$B=12$ triangular:}\\
The first stationary point of $|\mathcal{F}|$ at $q^2 \approx 2.3 \, \rm{fm}^{-2}$ does not have a corresponding stationary 
point for $\sqrt{\mathcal{I}}$ (Figure \ref{FF12T}). 
The next two stationary points are in good agreement, but
$\sqrt{\mathcal{I}}$ has a broad, flat maximum in-between.
\noindent
This difference can be explained via looking at the profiles of $\rho(r) r^2$ and 
$P(w) w^2$, recalling that $\rho(r)$ and $P(w)$ refer to the spherically averaged charge density and Patterson function 
respectively.

\begin{figure}[h]
\centering
\begin{minipage}{.5\textwidth}
  \centering
\includegraphics[scale=0.50,keepaspectratio=true]{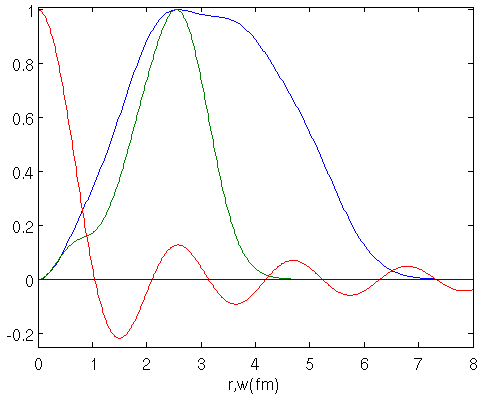}
  \captionof{figure}{Radial Profiles for the triangular\\ $B=12$ Skyrmion: $\rho(r) r^2$ (green), $P(w) w^2$ (blue) and 
  $j_0(qw)$ (red)}
  \label{RP12T}
\end{minipage}%
\begin{minipage}{.5\textwidth}
  \centering
\includegraphics[scale=0.50,keepaspectratio=true]{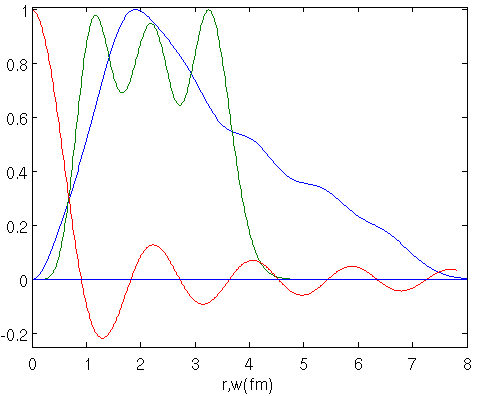}
  \captionof{figure}{Radial Profiles for the linear\\ $B=12$ Skyrmion: $\rho(r) r^2$ (green), $P(w) w^2$ (blue) and 
  $j_0(qw)$ (red)}
  \label{RP12L}
\end{minipage}
\end{figure}

Figure \ref{RP12T} displays radial profiles relevant to the triangular $B=12$ Skyrmion. 
The blue curve represents $P(w) w^2$, the green curve represents $\rho(r) r^2$ and the red curve 
represents $j_0(qw)=\frac{\sin qw}{qw}$, where $q^2 \approx 12 \, \rm{fm}^{-2}$ takes the value at the midpoint of the 
broad maximum of $\sqrt{\mathcal{I}}$.
As $q^2$ deviates slightly, away from $12 \, \rm{fm}^{-2}$, the third zero of $j_0(qw)$ moves across the maximum of $P(w) w^2$.
However, this deviation does not produce a large change in $\sqrt{\mathcal{I}}$ because $P(w) w^2$ has a broad maximum.
By contrast $\rho(r) r^2$ has quite a thin maximum, and this translates into a thinner maximum in $|\mathcal{F}|$.\\

\noindent
\underline{$B=12$ linear:}\\
There is better agreement between the location of zeros and stationary points of $|\mathcal{F}|$ and $\sqrt{\mathcal{I}}$ 
(Figure \ref{FF12L}) for the $B=12$ linear Skyrmion. 
The broad maximum of $\sqrt{\mathcal{I}}$ found for the triangular Skyrmion is no longer a feature because $P(w) w^2$ does 
not have a broad maximum.
The second zero of $|\mathcal{F}|$ is at a considerably larger value of $q^2$ than the corresponding zero for the triangular
Skyrmion.
This is equivalent to saying that $\mathcal{F}$ is negative for a larger range of $q^2$.
This increased range is caused by $\mathcal{F}$ having a negative maximum between the two zeros (at around 
$q^2 = 17 \, \rm{fm}^{-2}$); this feature can be explained by looking at the radial profile, $\rho(r) r^2$.
$\rho(r) r^2$ has three distinct peaks (Figure \ref{RP12L}), and the negative maximum of $\mathcal{F}$ occurs when the 
first non-trivial maximum of $j_0(qr)$ coincides with the central maximum of $\rho(r) r^2$.
$\mathcal{F}$ has a negative value because there is also coincidence of negative minima of $j_0(qr)$ 
with the two other positive maxima of $\rho(r) r^2$.

\subsubsection{Comparison of Intensity with Experimental Data for $B=12$ Skyrmions}
\vspace{1mm}

\begin{figure}[h]
\centering
 \includegraphics[scale=0.55]{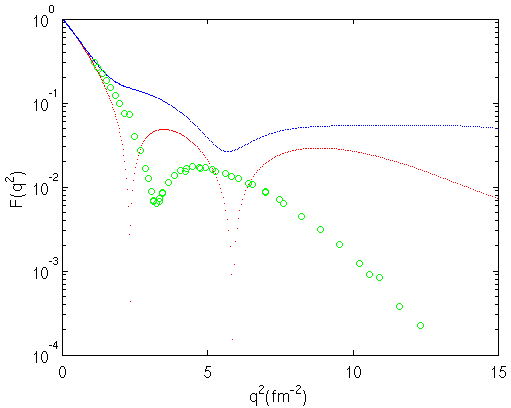}
  \captionof{figure}{Comparison of $\sqrt{\mathcal{I}}$ and $|\mathcal{F}|$ of the triangular $B=12$ Skyrmion with the Form\\ 
  Factor from Experimental Scattering Data for the $\rm{C}^{12}$ Nucleus}
  \label{D12}
\end{figure}

We only compare the triangular Skyrmion with the experimental scattering
data for $\rm{C}^{12}$ \cite{SM} because this Skyrmion corresponds
to the ground state of $\rm{C}^{12}$. (Figure \ref{D12})

The comparison of the scattering data with the classically averaged
intensity $\sqrt{\mathcal{I}}$ suggests that the shoulder found in 
$\sqrt{\mathcal{I}}$ around $q^2 = 2.5 \, \rm{fm}^{-2}$ is not a desirable feature. 
The classical averaging method does, however, agree with the 
data about the number of minima in the range of $q^2$ considered.

$|\mathcal{F}|$ has quite a good agreement with the data up to the first minimum. But it predicts a 
second minimum shortly after the first, and this is not found in the experimental data.

\subsection{$B=16$}
\vspace{2mm}

\begin{figure}[h]
\centering
\begin{minipage}{.5\textwidth}
  \centering
\includegraphics[scale=0.5,keepaspectratio=true]{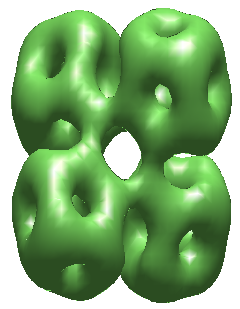}
  \captionof{figure}{Charge Density Isosurface\\ of the $B=16$ Skyrmion}
  \label{CD16}
\end{minipage}%
\begin{minipage}{.5\textwidth}
  \centering
\includegraphics[scale=0.5,keepaspectratio=true]{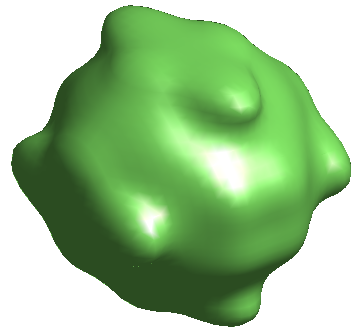}
  \captionof{figure}{Patterson Function Isosurface\\ of the $B=16$ Skyrmion}
  \label{PF16}
\end{minipage}
\end{figure}

\subsubsection{Patterson Function of the $B=16$ Skyrmion}
\vspace{1mm}

\begin{figure}[h]
\centering
\begin{minipage}{.5\textwidth}
  \centering
\includegraphics[scale=0.55,keepaspectratio=true]{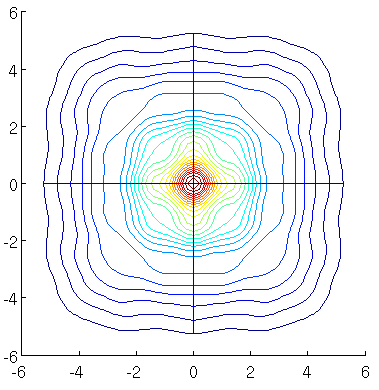}
  \captionof{figure}{Patterson Function Contours in\\ the $(x,z)$-plane for the $B=16$ Skyrmion}
  \label{PC161}
\end{minipage}%
\begin{minipage}{.5\textwidth}
  \centering
\includegraphics[scale=0.55,keepaspectratio=true]{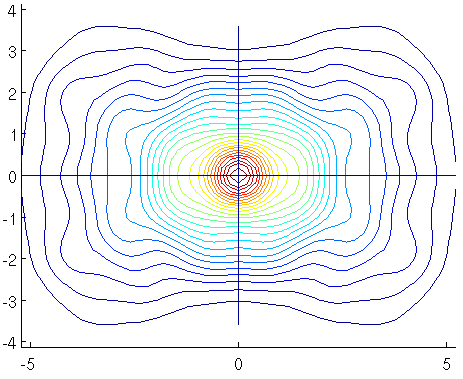}
  \captionof{figure}{Patterson Function Contours in\\ the $(x,y)$-plane for the $B=16$ Skyrmion}
  \label{PC162}
\end{minipage}
\end{figure}

The $B=16$ Skyrmion (Figure \ref{CD16}) is a 'bent square' lying in
the $(x,z)$-plane. It is midway between a flat square and a tetrahedrally-symmetric
arrangement of $B=4$ cubes \cite{BMS}. If one superimposes a second copy that has undergone an inversion, 
$\w \rightarrow -\w$, then this combination would have $D_{4h}$
symmetry about the $y$-axis. 
The Patterson function has an intrinsic inversion symmetry, so it has this $D_{4h}$ symmetry. 
Thus we have another example of the Patterson function having more symmetry than the Skyrmion that it is derived from.
The Patterson function isosurface (Figure \ref{PF16}) and contour plots (Figures \ref{PC161} and \ref{PC162}) display this
additional symmetry.

\subsubsection{Intensity of the $B=16$ Skyrmion}
\vspace{1mm}

\begin{figure}[h]
\centering
\begin{minipage}{.5\textwidth}
\centering
\includegraphics[scale=0.50]{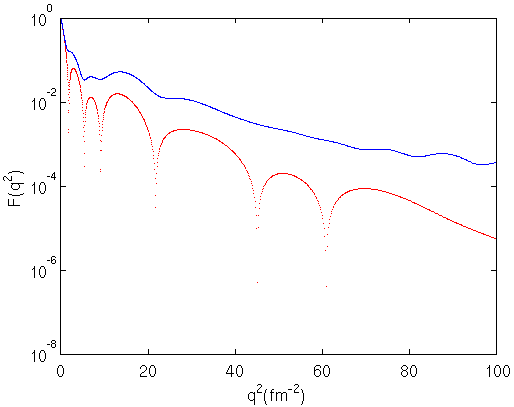}
  \captionof{figure}{$\sqrt{\mathcal{I}}$ and $|\mathcal{F}|$ for the $B=16$ Skyrmion}
  \label{FF16}
\end{minipage}%
\begin{minipage}{.5\textwidth}
\centering
 \includegraphics[scale=0.50]{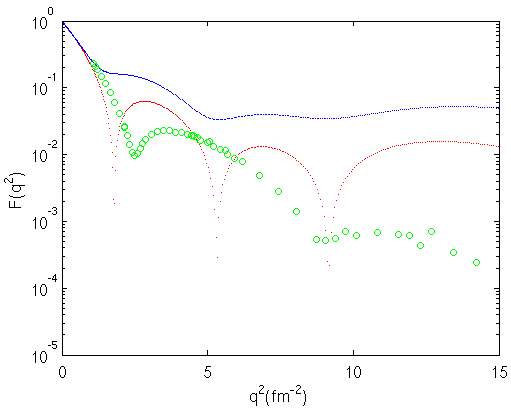}
  \captionof{figure}{Comparison of $\sqrt{\mathcal{I}}$ and
    $|\mathcal{F}|$ of the $B=16$ Skyrmion with the Form Factor from 
  Experimental Scattering Data for the $\rm{O}^{16}$ Nucleus}
  \label{D16}
\end{minipage}
\end{figure}

$\sqrt{\mathcal{I}}$ has a shoulder where $|\mathcal{F}|$ has a zero at around $q^2 = 1.8 \, \rm{fm}^{-2}$ (Figures \ref{FF16}
and \ref{D16}). This 
shoulder occurs when the first minimum of $j_0(qw)$  coincides with the maximum of $P(w) w^2$. 
$P(w) w^2$ has a broad maximum, which leads to a broad region of almost constant $\sqrt{\mathcal{I}}$.
After this initial difference there is good agreement between the locations of stationary points and zeros of 
$\sqrt{\mathcal{I}}$ and $|\mathcal{F}|$.

\subsubsection{Comparison of Intensity with Experimental Data for the $B=16$ Skyrmion}
\vspace{1mm}

The calculated intensities for the $B=16$ Skyrmion are compared with experimental electron scattering data off a $\rm{O}^{16}$ nucleus
\cite{SM} in Figure \ref{D16}. 
The colour scheme is the same as in Figures \ref{EXP4} and \ref{D12}.

We have further reinforcement that the shoulders observed in the 
classically averaged intensities of Skyrmions are not a desirable 
feature; they are again not present in the experimental data. The
classically averaged $\sqrt{\mathcal{I}}$ intensity bears little 
resemblance to the data curve.

The form factor $|\mathcal{F}|$ again produces a closer fit to the data with quite good agreement up to the first minimum
of the data curve. However, it still 
predicts too many minima in the data range that we are considering.

\subsection{$B=32$}
\vspace{2mm}

\subsubsection{Patterson Function of the $B=32$ Skyrmion}
\vspace{1mm}

\begin{figure}[h]
\centering
\begin{minipage}{.5\textwidth}
  \centering
\includegraphics[scale=0.45,keepaspectratio=true]{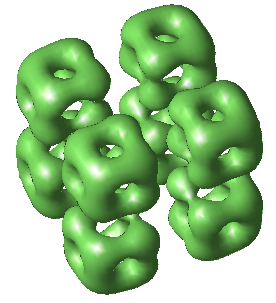}
  \captionof{figure}{Charge Density Isosurface\\ of the $B=32$ Skyrmion}
  \label{CD32}
\end{minipage}
\begin{minipage}{.5\textwidth}
  \centering
\includegraphics[scale=0.5,keepaspectratio=true]{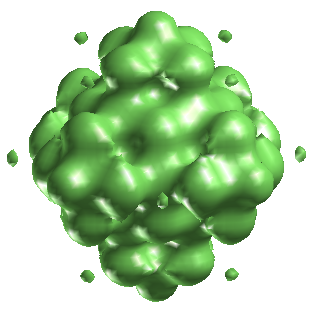}
  \captionof{figure}{Patterson Function Isosurface\\ of the $B=32$ Skyrmion}
  \label{PF32}
\end{minipage}
\end{figure}

The $B=32$ Skyrmion has cubic symmetry (Figure \ref{CD32}); it combines the internal symmetry of each $B=4$ cube with 
the cubic arrangement of these cubes. 
This translates into full cubic symmetry of the Patterson function. 
For small shift vectors $\w$ the structure of the Patterson function is similar to that of the isolated $B=4$ Skyrmion because 
the cubes are not yet overlapping. 

\noindent
As we increase the shift vector, the Patterson function has local maxima occurring at shift vectors going between 
the centres of cubes; these manifest themselves as disjoint balls in the isosurface plots (Figure \ref{PF32}). 
Interestingly there are now some local minima, which appear as disjoint balls lying within the volume enclosed by 
the main isosurface. These minima occur at shift vectors going between the corner of a cube
and the central cavity of a different cube. 

The contour plots (Figure \ref{PC32}) give further information;
$P(\w)$ is low along vertical or horizontal lines with an 
axis intercept of $2.6$ Skyrme units. 
$P(\w)$ is high for axis intercepts of $3.2$ Skyrme units. 
This shows that alignment of the cubes in just one axis direction
leads to a high value of $P$. 

\subsubsection{Intensity of the $B=32$ Skyrmion}
\vspace{1mm}

\begin{figure}[h]
\centering
\begin{minipage}{.5\textwidth}
  \centering
\includegraphics[scale=0.5,keepaspectratio=true]{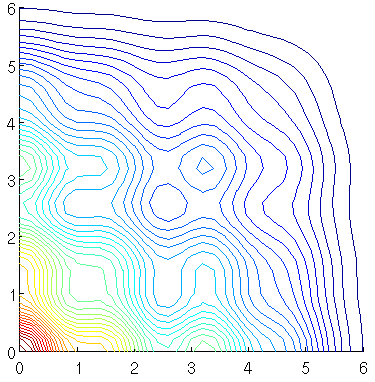}
  \captionof{figure}{Patterson Function Contours in\\ the $(x,y)$-plane for the $B=32$ Skyrmion}
  \label{PC32}
\end{minipage}%
\begin{minipage}{.5\textwidth}
  \centering
\includegraphics[scale=0.5]{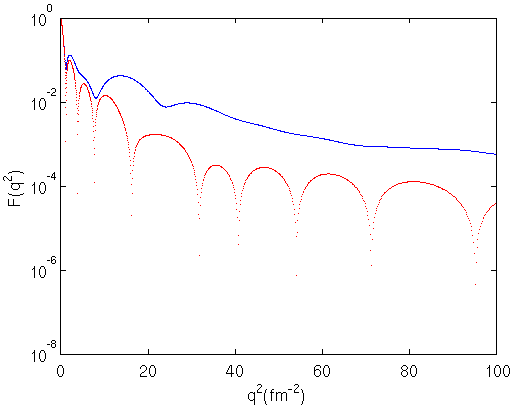}
  \captionof{figure}{$\sqrt{\mathcal{I}}$ and $|\mathcal{F}|$ for the\\ $B=32$ Skyrmion}
  \label{FF32}
\end{minipage}
\end{figure}

$\sqrt{\mathcal{I}}$ and $|\mathcal{F}|$ are in good agreement up 
to the first maximum at $q^2 \approx 2.2 \, \rm{fm}^{-2}$ (Figure \ref{FF32}).
As $q^2$ increases, the agreement gets worse because
$\sqrt{\mathcal{I}}$ has a shoulder
whereas $|\mathcal{F}|$ has a minimum/maximum pair, but both curves have a minimum 
around $q^2 = 7.7 \, \rm{fm}^{-2}$.

\subsection{$B=108$}
\vspace{2mm}
\begin{figure}[h]
\centering
\begin{minipage}{.5\textwidth}
  \centering
\includegraphics[scale=0.45,keepaspectratio=true]{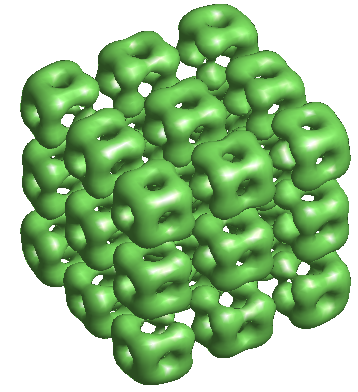}
  \captionof{figure}{Charge Density Isosurface\\ of the $B=108$ Skyrmion}
  \label{CD108}
\end{minipage}%
\begin{minipage}{.5\textwidth}
  \centering
\includegraphics[scale=0.5,keepaspectratio=true]{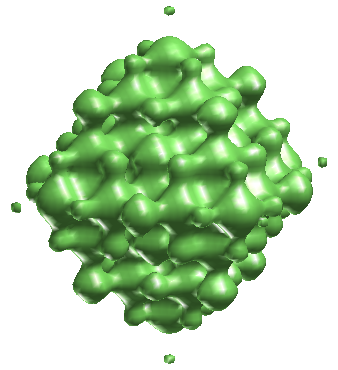}
  \captionof{figure}{Patterson Function Isosurface\\ of the $B=108$ Skyrmion}
  \label{PF108}
\end{minipage}
\end{figure}

The $B=108$ Skyrmion is a larger version of the $B=32$ Skyrmion \cite{FLM}. It is a cubic arrangement of $27$ $B=4$ cubes
(Figure \ref{CD108}). 

\subsubsection{Patterson Function of the $B=108$ Skyrmion}
\vspace{1mm}

\begin{figure}[h]
\centering
\begin{minipage}{.5\textwidth}
  \centering
\includegraphics[scale=0.55,keepaspectratio=true]{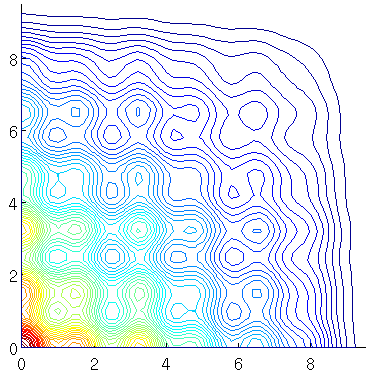}
  \captionof{figure}{Patterson Function Contours in \\the $(x,y)$-plane for the $B=108$ Skyrmion}
  \label{PC108}
\end{minipage}%
\begin{minipage}{.5\textwidth}
  \centering
\includegraphics[scale=0.50]{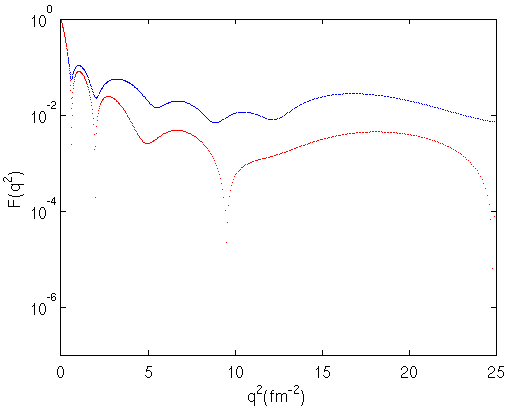}
  \captionof{figure}{$\sqrt{\mathcal{I}}$ and $|\mathcal{F}|$ for the\\ $B=108$ Skyrmion}
  \label{FF108}
\end{minipage}
\end{figure}

The Patterson function of the $B=108$ Skyrmion (Figure \ref{PF108}) is 
similar to that of the $B=32$ Skyrmion, only more 
detailed. Another layer of cubes in the Skyrmion
increases the number of shift vectors going between $B=4$
cube corners, which causes more stationary points of $P(\w)$, as seen in Figure \ref{PC108}.

\subsubsection{Intensity of the $B=108$ Skyrmion}
\vspace{1mm}

The intensity (Figure \ref{FF108}) is also rather similar to that 
of the $B=32$ Skyrmion; there are now more stationary points 
due to an increased number of relevant shift vectors.

\subsection{The Skyrme Crystal}
\vspace{2mm}

\begin{figure}[h]
\centering
 \includegraphics[scale=0.55]{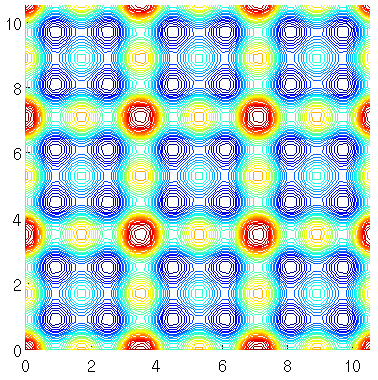}
  \captionof{figure}{Patterson Function Contours in \\the $(x,y)$-plane for the Skyrme Crystal}
\label{Crystal}
  \end{figure}

The Skyrme crystal can be viewed as an infinite cubic lattice 
of $B=4$ Skyrmions with enhanced symmetries \cite{MS}. 
Its Patterson function can be well approximated by taking the 
centre cube from a $B=108$ Skyrmion, creating a 
$3\times3\times3$ grid of these, and calculating the Patterson
function for the centre cube of this grid shifting towards its 
neighbours. The result is then extended periodically.
This periodic Patterson function (Figure \ref{Crystal}) can be used in
reverse to make a prediction for the Patterson function of the $B=108$ Skyrmion.
When the $B=108$ Skyrmion is shifted, some of its cubes move into the 
empty space surrounding the Skyrmion, and these cubes do not contribute.
This can be accounted for by a multiplicative factor, and when this is
included, one gets a close fit 
to the actual Patterson function of the $B=108$ Skyrmion.

\section{Approximations for the Zeros and Stationary Points of $\mathcal{F}$}
\vspace{3mm}

It is important to know the locations of the zeros and stationary points of the form factor $\mathcal{F}$.
These locations are related to intrinsic features of the charge
density, and it is useful to have an approximate method for finding them.

\subsection{Zeros of the Form Factor}
\vspace{2mm}

Consider the expression for the form factor $\mathcal{F}$ in terms of
the spherically averaged charge density $\rho(r)$,
\begin{equation}
\mathcal{F}(q^2)=\int \rho(r) \,
e^{-i \q \cdot \x} \, d^3\x = 4\pi \int_0^{\infty}
\rho(r) \, \frac{\sin{qr}}{qr} r^2 \, dr \,.
\end{equation}
Looking at this expression, it is not immediately obvious where
the zeros of $\mathcal{F}$ are. 
However, $\rho(r) \, r^2$ is quite localised and approximately symmetric about its maximum, so a truncated Taylor
expansion of $\frac{\sin{qr}}{qr}$ about its first zero at $qr = \pi$ 
should lead to a good approximation to a value of $q^2$ where 
$\mathcal{F}$ is zero.

For $qr$ close to $\pi$,
\begin{equation}
\frac{\sin{qr}}{qr}\simeq-\frac{1}{\pi}(qr-\pi)+\frac{1}{\pi^2}(qr-\pi)^2 \,,
\end{equation}
\noindent
which gives an estimate for a zero of $\mathcal{F}$,
\begin{equation}
\mathcal{F}(q^2) \simeq -4\pi \int_0^{\infty} \rho(r) \, r^2 \left(\, \frac{1}{\pi}(qr-\pi)-\frac{1}{\pi^2}(qr-\pi)^2 \, \right) 
\, dr=0 \,.
\end{equation}
This implies that 
\begin{equation}
\langle r^2 \rangle \left(\frac{q}{\pi}\right)^2 -3 \langle r \rangle \left(\frac{q}{\pi}\right) + 
2= 0 \,,
\label{Quad}
\end{equation}
where 
\begin{equation}
\langle r^n \rangle = \int \rho(\x) \, r^n \, d^3 \x 
= 4\pi \int_0^{\infty} \rho(r) \, r^{n+2} \, dr \,.
\end{equation}
The solutions are
\begin{equation}
\frac{q}{\pi}=\frac{3\langle r \rangle \pm \sqrt{9\langle r \rangle^2-8\langle r^2 \rangle}}{2\langle r^2 \rangle} \,,
\end{equation}
and we choose the lower sign (smaller $q$) because the upper sign corresponds to the second zero of $\frac{\sin{qr}}{qr}$, 
and the approximation is poor to the right of this zero. We also only want to include one zero in the region of non-zero 
charge density.
In conclusion, one can make a prediction for the location of the first zero of $\mathcal{F}$ using only the first few 
radial moments of the charge density.

\subsection{Stationary Points of the Form Factor}
\vspace{2mm}

The expression for $\frac{d\mathcal{F}}{dq^2}$ in terms of the charge
density is
\begin{equation}
\frac{d\mathcal{F}}{dq^2}= \frac{1}{2q^2}\int \rho(r) r^2\left(\cos{qr}-\frac{\sin{qr}}{qr}\right)\, dr \,.
\end{equation}
Using a truncated Taylor series of $\left(\cos{qr}-\frac{\sin{qr}}{qr}\right)$ about its first positive zero leads
to good approximation of the location of zeros of $\frac{d\mathcal{F}}{dq^2}$.
The location of the first zero of $\left(\cos{qr}-\frac{\sin{qr}}{qr}\right)$ will be labeled by $qr=b$.
The quadratic term of the Taylor series helpfully vanishes, so we are left with a linear equation to solve for $q$
\begin{equation}
\frac{d\mathcal{F}}{dq^2}\simeq -\frac{1}{2q^2}\int \rho(r) \sin{b} \, (qr-b) \, dr,
\end{equation}
which implies that
\begin{equation}
 q= \frac{b}{\langle r \rangle} .
 \label{Lin}
\end{equation}
\noindent
The stationary point lies at a value of $q^2$ such that $q\langle r \rangle$ ($q$ times the mean of 
$\rho(r) r^2 $) is equal to $b$.

Inspecting the values of $q^2$ that give the first minima of $\mathcal{I}$ rather than $\mathcal{F}$, it would appear 
that we would have to consider the series expansion about the first three zeros (excluding $qr=0$) of 
$\cos{qr}-\frac{\sin{qr}}{qr}$, in order to approximate the location of the first minimum.
A poor approximation can be found by scaling $q^2$ such that the second zero of $\cos{qw}-\frac{\sin{qw}}{qw}$ is close to the 
the maximum of $P(w) w^2$.

Many Skyrmions have good agreement between the locations of stationary points of $\sqrt{\mathcal{I}}$ and those of 
$\mathcal{F}$, but there does not seem to be a strong analytical link showing why. 
Some stationary points are only present in either $\sqrt{\mathcal{I}}$ or $\mathcal{F}$, 
so this good agreement cannot be used to approximate the locations of the stationary points of $\sqrt{\mathcal{I}}$ 
based on the approximations for $\mathcal{F}$.

\subsection{Accuracy of Approximations}
\vspace{2mm}

Table \ref{T1} presents our estimated values of $q^2$ at the first
zero and first minimum of $\mathcal{F}$ compared to
the actual values.

\begin{table} [h]
\centering
    \begin{tabular}{| l | c | c | c | c |}
    \hline
    Baryon & Approx. location  & Actual location & Approx. location & Actual location \\ 
    Number & of first zero  & of first zero & of first minimum & of first minimum \\ \hline
    $B=4$ & $4.57$ & $4.57$ & $7.47$ & $8.01$ \\ \hline
    $B=8$ twisted & $2.37$ & $2.37$ & $3.53$ & $3.65$ \\ \hline
    $B=12$ triangular & $2.38$ & $2.31$ & $3.66$ & $3.50$ \\ \hline
    $B=12$ linear & n/a & $8.18$ & $3.89$ & $11.63$ \\ \hline
    $B=16$ & $1.80$ & $1.78$ & $2.93$ & $2.82$ \\ \hline
    $B=32$ & $1.15$ & $1.15$ & $2.04$ & $1.96$ \\ \hline
    $B=108$ & $0.61$ & $0.62$ & $1.07$ & $1.02$ \\ 
    \hline
    \end{tabular}
    \caption{Approximate and actual locations of first
      zero and first minimum of $\mathcal{F}(q^2)$}
    \label{T1}
    \end{table}  

The table shows that the approximations of the first zero and minimum of $\mathcal{F}$ are generally very good. 
The exception is the linear $B=12$ Skyrmion; the quadratic equation (\ref{Quad}) for the 
zero of $\mathcal{F}$ has no roots, and the prediction (\ref{Lin}) for the minimum is very far away from the actual location. 
This can be explained by noting the presence of the shoulder near $q^2 = 4 \, \rm{fm}^{-2}$ of $\mathcal{F}$ 
(Figure \ref{FF12L}) for the linear $B=12$ Skyrmion. 
The location of the shoulder is close to our predicted location for the first stationary point of $\mathcal{F}$.
The approximation finds the shoulder rather than the first true minimum of $\mathcal{F}$.
The fact that $\mathcal{F}$ does not have a zero before the shoulder explains why the quadratic equation (\ref{Quad})
has no roots in this case.

\section{Conclusions}

We have calculated the charge density and Patterson function for several Skyrmions with baryon number a multiple of four.
From these we have calculated the quantum averaged electron scattering intensity $|\mathcal{F}(q^2)|^2$ and the classically 
averaged electron scattering intensity $\mathcal{I}(q^2)$ for the Skyrmion's states with spin and isospin zero and found
that there is a sizable difference between them. 
This shows that it is important to make the distinction when considering non-spherically symmetric charge distributions. 
However, we have found that neither method gives results in good agreement with experimental electron scattering data.

The quantum averaged $|\mathcal{F}|$ gives the better fit to the experimental data; however, there are still some problems.
For all Skyrmions that we have considered, the calculated intensity has too many minima when compared to the scattering data 
and the entire curve $|\mathcal{F}(q^2)|$ generally lies above the data points (a problem exacerbated for the classically
averaged intensity because of the Cauchy inequality).

There is also the problem that the calculated first minimum of the intensity is at too low a momentum and that the intensity
at the second maximum has a higher value than the experimental data. We have investigated the effect of introducing a 
central depression for a few well known charge distributions and found that a large depression causes the various problems 
stated above.
This would indicate that the region of near-zero baryon density found at the centre of many of the Skyrmions is a feature which 
does not agree with experimental electron scattering data.

However, we have treated the Skyrmion as a rigid body in this paper and have not taken into account the vibrational modes 
of the Skyrmion. 
If we had included these modes, then the ground state should be considered as an average over configurations with different baryon densities 
which lie within the vibrational space.
This vibrational averaging would result in a charge distribution with significantly less structure because the central 
region of near-zero baryon density would be partially washed-out.
The averaging would still result in a slight central depression but this is believed to be a feature of nuclear charge 
distributions, particularly for larger nuclei \cite{H}.

We have found that it is possible to predict the locations of the first zero and stationary point of $|\mathcal{F}|$ using 
moments of the charge density.
Disappointingly, we have also found that it is difficult to predict the locations of stationary points of 
$\sqrt{\mathcal{I}}$ using moments of the Patterson function. 
However, given their similarities, there is probably a link between the stationary points of $\sqrt{\mathcal{I}}$ and 
$|\mathcal{F}|$.

\section*{Acknowledgements}

C. King is supported by an STFC studentship.

\end{document}